Irvan B. Arief-Ang, Flora D. Salim, and Margaret Hamilton

# CD-HOC: Indoor Human Occupancy Counting using Carbon Dioxide Sensor Data

Irvan B. Arief-Ang, Flora D. Salim, Margaret Hamilton*

*School of Computer Science and Information Technology, RMIT University, Melbourne, Australia*

**Abstract**

Human occupancy information is crucial for any modern Building Management System (BMS). Implementing pervasive sensing and leveraging Carbon Dioxide data from BMS sensor, we present Carbon Dioxide - Human Occupancy Counter (CD-HOC), a novel way to estimate the number of people within a closed space from a single carbon dioxide sensor. CD-HOC de-noises and pre-processes the carbon dioxide data. We utilise both seasonal-trend decomposition based on Loess and seasonal-trend decomposition with moving average to factorise carbon dioxide data. For each trend, seasonal and irregular component, we model different regression algorithms to predict each respective human occupancy component value. We propose a zero pattern adjustment model to increase the accuracy and finally, we use additive decomposition to reconstruct the prediction value. We run our model in two different locations that have different contexts. The first location is an academic staff room and the second is a cinema theatre. Our results show an average of 4.33% increment in accuracy for the small room with 94.68% indoor human occupancy counting and 8.46% increase for the cinema theatre in comparison to the accuracy of the baseline method, support vector regression.

## 1. Introduction

Data obtained from the U.S. Department of Energy indicates that 35% - 45% of total maintenance costs within a building are for heating, ventilation, and air conditioning (HVAC) [14]. Hence, there are substantial investments in the energy usage research area to automate HVAC control in buildings based on occupancy patterns. Reducing HVAC usage will massively lessen overall energy consumption. However, this may compromise the comfort of the dwellers. A Building Management System (BMS) can intelligently adjust the HVAC based on the occupancy pattern.

The majority of buildings, especially older ones, do not have an adequate infrastructure to sense people and where they are within a building accurately. Hence it is challenging to determine the precise ground truth value for analysis purposes. Previous research mainly focussed on the use of simulation models [22, 39] to reduce energy consumption. Methods of simulating the occupants' behaviour with the aim of reducing energy consumption based on their behaviours were proposed in [41, 42]. Unfortunately, simulations with agentbased models do not reflect the behaviour and uncertainty from the actual environment and are not adaptable to different types of rooms and buildings. Therefore, the better approach is to analyse the data collected from the real world.

*Email addresses: irvan.ariefang@rmit.edu.au, flora.salim@rmit.edu.au, margaret.hamilton@rmit.edu.au

Using sensor data for indoor human occupancy detection is the current trend in ambient sensing research area [3, 8, 16, 25, 29, 38, 48]. In [3], it was highlighted that carbon dioxide ($CO_2$) is the best ambient sensor predictor for detecting human presence. By using only $CO_2$, 91% accuracy was achieved for binary prediction, knowing the room is occupied or vacant [5] and have 15% accuracy for recognising the number of occupants. A hidden Markov model (HMM) was implemented for $CO_2$ dataset to predict human occupancy and 65% - 80% range of accuracy was achieved for predicting up to 4 occupants [32].

There are two main motivations for this research. The first motivation is to help space and room utilisation. By knowing the number of people in each room at a given time, the building manager can monitor which rooms are under-utilised and which rooms are over-utilised. With this knowledge, he can adjust room allocation and utilisation accordingly. The second motivation is to support BMS so it can reduce their power consumption when there is no person in the room. Knowing the number of persons at a given time for each room becomes crucial to achieve this motivation.

In this paper, we propose carbon dioxide - human occupancy counter (CD-HOC), a new method to count indoor human occupancy based on the amount of $CO_2$ in the air. The main reason we use $CO_2$ data is because $CO_2$ sensors are often installed in modern BMSs and one sample of $CO_2$ data regression chart is shown in Figure 1.

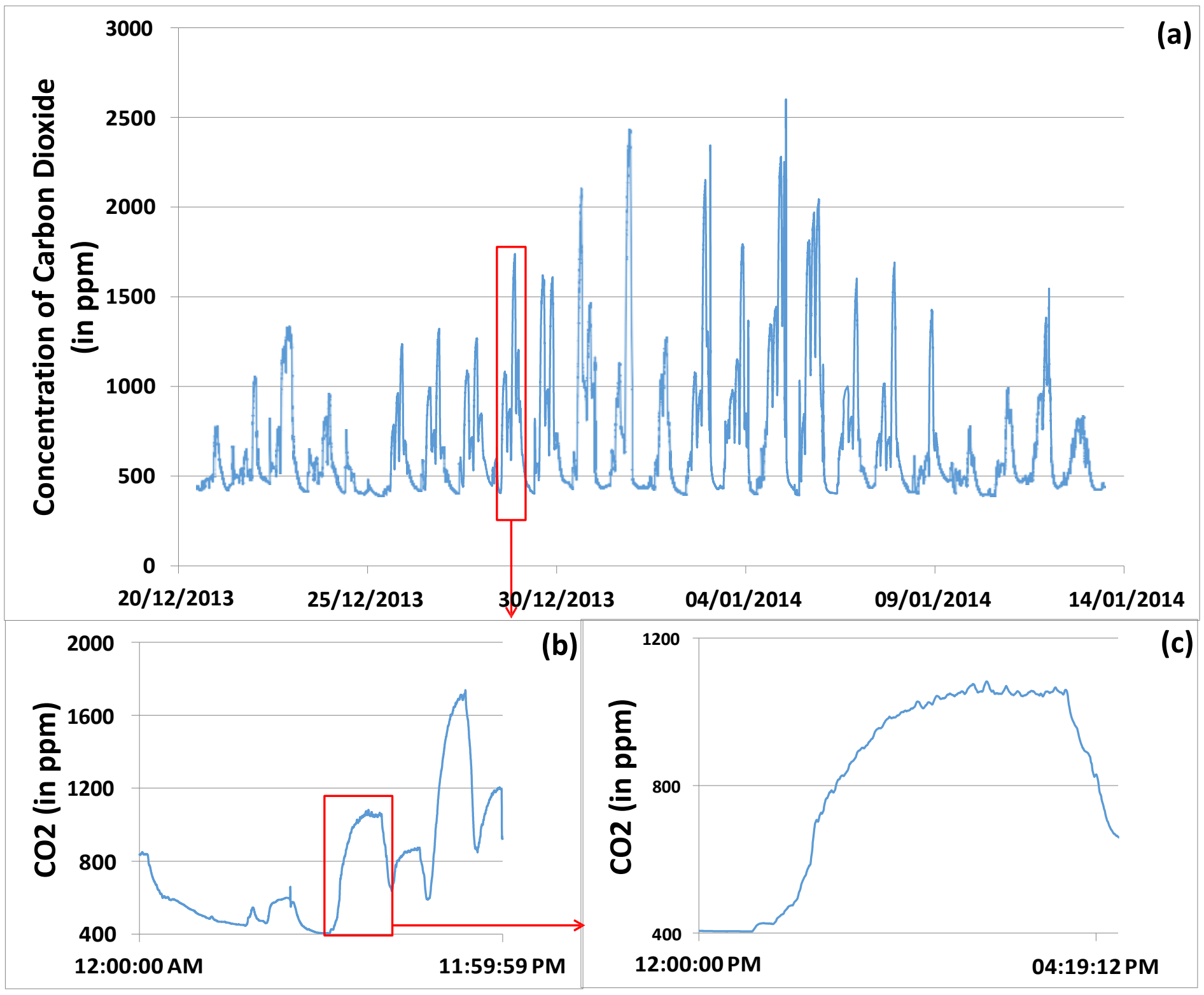

Figure 1: $CO_2$ concentration in the movie theatre over the time. Figure 1a shows an overview of all measurements. Days, times and movies screening can be easily identified. Figure 1b shows the concentration on December 28th. Movie screening can be seen and the shape of each screening can be identified differently. Figure 1c shows the $CO_2$ concentration during the movie "The Hunger Games 2: Catching Fire" on December, 28th 2013 at 13:15.

As humans exhale $CO_2$ while they breathe, there is a correlation between $CO_2$ and occupancy. Machine learning model experiments were performed on two different sized rooms with the capacity up to 300 occupants. Three advantages of this method are: a) CD-HOC ensures that users' privacy is protected; b) it employs low equipment cost due to pre-installation; and c) it only uses $CO_2$ data, reducing the chance of errors caused by data integration.

### 1.1. Research Contribution

The main contributions of this paper are as follows:

- We propose a novel feature engineering method to process both $CO_2$ and human occupancy data into four main features: trend, seasonal, irregular and zero pattern adjustment.
- We develop a time-series model to predict the number of humans occupying a closed space. The accuracy of the proposed method is higher than the current state-of-the-art machine learning algorithm method known as support vector regression (SVR).
- We compare our proposed two customised seasonal decomposition methods, to solve the human occupancy prediction problem and highlight the advantages of each method.
- We generalise our method to be implemented for any prediction problem $Y_n$ if $X_n$ is known and $Y_n$ is dependent to $X_n$ where n is the number of sample points.

The remainder of the paper is organised as follows. Section 2 presents the related work on current state- of-the-art indoor human occupancy methods. Section 3 covers the problem definition. Section 4 introduces a new carbon dioxide occupancy counter model. Section 5 describes experiments, results, comparisons with the state-of-the-art algorithm and discussion. Finally, section 6 concludes the paper with directions to the future work.

## 2. Background and Related Work

Occupancy detection has been explored in the last decade and one of the biggest challenges is to do this without using image processing from cameras. When using image processing techniques [18, 33], the levels of accuracy for human occupancy detection can reach up to 80%. Unfortunately, this method raises privacy concerns. Research communities have been doing their best to propose various methods to detect human occupancy without using cameras or image processing. Due to the extensive method of machine learning algorithms that have been used for human occupancy recognition, a machine learning terms list methods is listed presented Table A.1 in the Appendix.

In this section, we divide related works of human occupancy research into six subsections based on their method. Research about human occupancy by simulation is presented in Section 2.2, by radio-based in Section 2.3, by singular sensor-based in Section 2.4, by multi sensor-based in Section 2.5 and the ones that closely relate to this paper using $CO_2$ sensor in Section 2.6. In Section 2.7, we present our summary of the current state of the human occupancy research domain.

### 2.1. Background Study of Human Occupancy Calculation

This subsection discusses human occupancy calculation with flow rate of $CO_2$ using metabolic rates. A person exhales $CO_2$ as the natural process of breathing, but the amount varies from person to person. Equation 1 below gives the exhalation rate equation of $CO_2$, derived from the metabolic rates formula [4].

$$V_{CO_2} = \frac{M \cdot RQ \sqrt{H \cdot W}}{21132 \cdot (0.23RQ + 0.77)} \quad (1)$$

| | |
|---|---|
| $V_{CO_2}$ | flow rate of $CO_2$ |
| M | metabolic rates (in $W/m^2$) |
| RQ | respiratory quotient |
| H | height (in $cm$) |
| W | weight (in $kg$) |

We note a non-linear relationship between the flow rate of $CO_2$ and the average value of person's height ($H$) and weight ($W$). Due to this fact, the non-linear relationship issue needs to be considered if we wish to correlate human feature (height and weight) with $CO_2$ data only. In our methodology, we have addressed this non-linear gap in the literature and will be explained further in the Section 4.3.

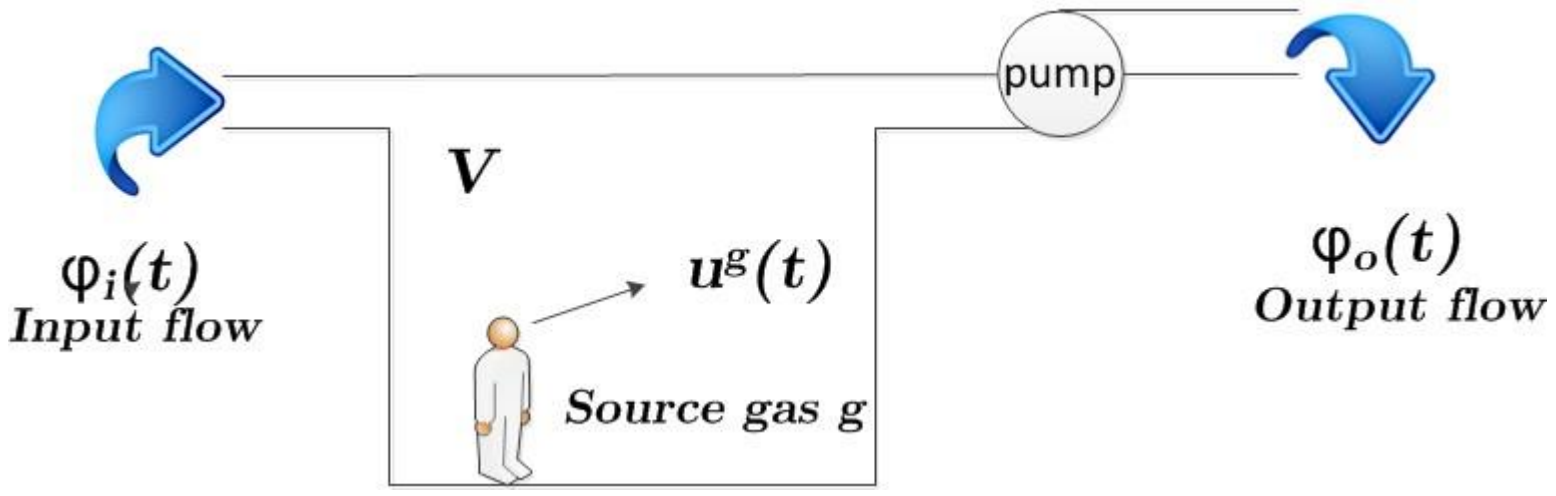

Figure 2: Simplified presentation of gas exchange in the respiratory chamber where $\phi i(t)$ is the flow rate of input air at time t, $\phi o(t)$ is the flow rate of output air at time t, V is the volume at time t, $u^g(t)$ is the gas production rate at time t.

The Size of the room is the next aspect that we need to consider to ensure our data is useful for indoor human occupancy prediction. Figure 2 explains the flow rate of inhalation [23]. For any given room that has standard ventilation, there will be an input flow $\phi i(t)$ and pump out by a specific output flow $\phi o(t)$. For specific room size $V$ at the time $t$, the gas production rate is $u^g(t)$. From this finding, the size of the room is crucial for understanding the gas production rate.

From Equation 1 and Figure 2, to be able to have an accurate indoor human occupancy by using metabolic rates formula and flow rate of $CO_2$, we can see that many parameters are required. Parameters include the occupant's weight, height, the rate of breath and also the size of the room. The need for occupant's weight and height can lead to privacy issues. In addition, it is very difficult to obtain accurate readings of room size.

### 2.2. Simulation based Indoor Human Occupancy Detection

The common method for detecting human occupancy is to construct a simulation model for a room or building [22, 39, 41, 42]. Simulation models will not be accurate for multiple buildings with different features and characteristics as there are many variables including the ones that mentioned at subsection 2.1 that need to be considered for indoor human occupancy.

Table 1: Models, parameters and reported accuracies for radio-based occupancy detection research.

| Source | Occupancy Detection Algorithms | Detailed Devices | Max. # of people | Duration | Location | Accuracy (Occupancy) |
|---|---|---|---|---|---|---|
| [12] | Multipath fading (MP) with line of sight, Kullback-Leibler (KL) divergence | WiFi transmitter and receiver | 9 | unknown | indoor and outdoor | 96% for outdoor and, 63% for indoor |
| [15] | Doppler equation, Neyman-Person detector (noise detector), Exponentially-weighted moving average (EWMA) | Ultrawideband radar | binary occupancy | unknown | outdoor | unknown |
| [34] | Pareto distribution of occupancy's inactivity | Microwave motion sensor | unknown | 10 days | staff office room | unknown |
| [26] | Support Vector Regression (SVR) | A low-power pulser radar | 40 | 750 minutes | four rooms | unknown |
| [35] | Neural Network (NN), SVM and Sequential Counting (SC) | WiFi Access Point, Samsung Galaxy S3 | 50 | five hours | two classrooms | 93% |
| [44] | K-nearest neighbours (KNN) | Radio Frequency (wireless signal) | 3 | unknown | corridor | unknown |

### 2.3. Radio-based Indoor Human Occupancy Detection

Radio-based devices include but are not limited to WiFi, Bluetooth, any electromagnetic waves and gamma rays. A summary of the radio based research is in Table 1.

A model using only WiFi power based on its received signal strength indication (RSSI) is used for occupancy detection [12]. By using the line of sight analysis, outdoor human occupancy accuracy can be up to 96%.

A low-power pulsed radar was utilised for people counting in [26]. The counting is modelled with support vector regression (SVR) for up to 40 occupants. A correlation coefficient of 0.97 was achieved with 2.17 mean absolute error between the estimated count and the ground truth.

Using wireless signals, device-free occupancy recognition is achieved [44]. They managed to distinguish the state of wireless signal between an empty environment and occupied one. Furthermore, this study successfully detected the occupants' activity such as walking, lying, crawling or standing.

In [34], a microwave motion sensor is used to detect motion and then the lighting is controlled with time delays to reduce electricity consumption. Duta et. al. take advantage of ultra wideband radar that can function beyond the line of sight to detect people [15]. However such equipment is expensive and rarely justified for the purpose of occupancy detection alone.

Table 2: Models, parameters and reported accuracies for single type of sensor-based occupancy detection research.

| Source | Occupancy Detection Algorithms | Detailed Devices | Max. # of people | Duration | Location | Accuracy (Occupancy) |
|---|---|---|---|---|---|---|
| [1] | Using hardware (CC2530 micro controller) build in capability | Reed swithes and PIR sensors | binary occupancy | 2 weeks | 10 offices | unknown |
| [13] | Bayesian probability theory and Belief Network (BN) | 3 PIR sensors and telephones sensor | binary occupancy | 2 days | 2 offices | 1 PIR: 20.1%<br>2 PIR: 97.8%<br>3 PIR: 99.9% |
| [20] | KNN and learning-based model predictive control (LBMPC) | Microsoft Kinect, infrared sensor and laser pointer | binary . occupancy | more than 3 months | office room | unknown |
| [21] | Motion detection | PIR sensors | binary occupancy | unknown | office room | unknown |
| [27] | Modified Bayesian combined forecasting approach with seasonal ARIMA model, historic average, time delay NN and SVR | 50 PIR sensors | unknown | 2 years | 2 large buildings | unknown |
| [46] | non-homogeneous Poisson model with two different exponential distributions | Infrared sensor behind a fresnel lens | unknown | 1 year | 35 single person offices | unknown |
| [5] | Non-negative matrix factorization (NMF), Ensemble Least Square Regression (ELSR) and SVR | $CO_2$ sensors | 15 (lab)<br>42 (classroom) | 13 days | a lab and a classroom | 91% for vacant prediction and 15% for occupied prediction |
| [36] | Classical Linear Minimum Variance (LMV) estimator | binary motion sensors | 1 | unknown | one room | unknown |
| [32] | Hidden Markov Models (HMM), NN and Support Vector Machines (SVM) Latent | $CO_2$ for both inside and outside room | 4 | 58 days | an open plan office with 16 rooms and one conference room | 65% - 80% |
| [9] | Latent dirichlet allocation (LDA) | PIR sensors | unknown | 24 weeks | 3 floor Innotek building | unknown |
| [6] | KNN, Linear regression (LR) and Artificial neural networks (ANN) | PIR and thermal array sensors | unknown | 3 weeks | a 17-node deployment covering 10 building areas | unknown |

### *2.4. Indoor Human Occupancy Detection with Homogeneous Sensors*

Indoor human occupancy detection with homogeneous sensors refers to research using one type of sensors for person counting. A summary of the homogeneous type of sensor based related work can be found in Table 2.

Algorithms to detect indoor human occupancy for the purpose of reducing energy cost are proposed in [1, 20, 21, 37, 6]. Energy saving results vary from 5% up to 60%.

The most common device that is used for person counting is a passive infra-red (PIR) sensor which is used as a motion sensor with a reed switch as the door sensor [1, 9, 13, 20, 21, 27, 36, 37, 46]. In [13], single, double and triple PIR sensors are compared to detect the presence of humans and the accuracy is 20.1%, 97.8% and 99.9% respectively.

Using sensor network data, 50 PIR sensors were deployed and the data was utilised to build a modified Bayesian forecasting method [27]. They compared the accuracy with several other machine learning methods such as seasonal autoregressive integrated moving average (ARIMA), neural networks (NN) and SVR with their techniques (modified Bayesian combined forecasting) and mentioned that their techniques have the lowest error. Their accuracy results were not presented.

ThermoSense, a system for estimating occupancy by using thermal array sensor combined with passive infrared sensor was presented in [6]. Using k-nearest neighbour (KNN), linear regression (LR) and artificial neural network (ANN), ThermoSense could predict the room occupancy and reduce energy saving up to 25%.

In [31], electricity consumption data from electricity meters was used as the feature in occupancy analysis. The article presents four models of the regression in SVM, KNN, thresholding (THR) and HMM.

Table 3: Models, parameters and reported accuracies for multi types of sensor-based occupancy detection research.

| Source | Occupancy Detection Algorithms | Detailed Devices | Max. # of people | Duration | Location | Accuracy (Occupancy) |
|---|---|---|---|---|---|---|
| [38] | Rule-based heuristic, Multi-Layer Perceptron (MLP), Gaussian processes (GP), LR, v-SVM-R and Ensemble Voting (EV) | BLEMS sensors | 16 | several weeks | 2 shared labs space | 46% - 95% |
| [45] | Density-based spatial clustering of applications with noise (DB-SCAN) and Maximum Likelihood Estimate (MLE) | ultrasonic distance sensor for height on the doorway, motion sensors and magnetic reed swith sensors | 20 (lab) 4 (home residents) | 5 days | 1 lab and 3 homes | 95% (identification accuracy) |
| [48] | SVM, ANN with Radial Basis Function (RBF) and HMM | light, sound, motion, $CO_2$, temperature, relative humidity and PIR sensors | 9 | 20 days | 2 shared labs space | 87.62% for self estimation and 64.83% for cross-estimation |
| [8] | Random Forest (RF), Gradient Boosting Machines (GBM), Linear Discriminant (LD) analysis and Classification and Regression Trees (CART) | Rasberry Pi with light, $CO_2$, DHT22 (temp/humid) sensors, Zigbee radio and digital video camera | 2 | 2 days | a room | 95% - 99% with all predictors and 83% - 85% by using temperature only |
| [30] | RF | acoustic (microphone), locomotive (accelerometer) and location (magnetometer) sensors | 8 | unknown | unknown | unknown |
| [29] | KNN and SVM | motion PIR, acoustic noise, temperature, light and humidity sensors | 20 | 14 days for low traffic area and 10 days for high traffic area | a large commercial buildings | unknown |
| [25] | Decision Trees (DT) | $CO_2$, computer current, light, PIR and sound sensors | 1 | 7 days | a single cubicle | $CO_2$: 94.68% Current: 96.27% Light: 81.02% Motion: 98.44% Sound: 90.79% |
| [16] | ANN with MATLAB and WEKA | $CO_2$, sound, relative humidity, temperature (air and computer) and PIR sensors | 39 | 7 days | an open-plan office | 84.59% |
| [17] | ANN with MATLAB and WEKA | sound, case temperature, humidity, light, $CO_2$, motion, Volatile Organic Compounds (VOCs) and PIR sensors | 6 | 30 days | an open-plan office | 75% |
| [49] | ANN, DT, KNN, Naïve Bayesian (NB), Tree augmented Naïve Bayes Network,(TAN) and SVM | light, sound, PIR, $CO_2$, Reed door sensor, relative humidity and temperature sensors | 3 (single occupancy rooms) 9 (multi occupancy rooms) | 1 month (single occupancy rooms) 20 days (multi occupancy rooms) | Two single-occupancy rooms and two multi-occupancy rooms | ANN: 92.5%-97.1% DT: 96.0%-98.2% KNN: 95.4%-97.5% NB: 88.9%-94.3% TAN: 95.3%-98.0% SVM: 95.1%-97.5% |
| [28] | unknown | 64 wired BMS sensors, 50 moveable sensor boxes and several cameras | 8 | 2 weeks | 7 rooms | unknown |
| [3] | MLP, GP with RBF, SVM, RF and NB | temperature, humidity, $CO_2$, sound, pressure and illumination (light) sensors | binary occupancy | 2 weeks | one single person office room | 96% - 99% |

### 2.5. Indoor Human Occupancy Detection with Heterogeneous Sensors

Indoor human occupancy detection with heterogeneous sensors refers to any research that uses multiple types of sensors for person counting. Table 3 presents a summary of related work on indoor occupancy detection using heterogeneous sensors.

The Building Level Energy Management Systems (BLEMS) project from University of Southern California uses a combination of sensors (light, sound, motion, $CO_2$, temperature and humidity sensor) to create a model to estimate the human occupancy. A radial basis function (RBF) method was used to find 87.62% for self-estimation and when the model was implemented in another room, the cross-examination result showed 64.83% occupancy accuracy [48]. Various machine learning algorithms such as multi-layer perceptron (MLP), Gaussian processes (GP), LR, SVM and ensemble voting (EV) were implemented to find a range of occupancy from 46% - 95% [38].

Utilising density-based spatial clustering of applications with noise (DB-SCAN), [45] produced 95% accuracy with ultrasonic distance sensors for height sensors, motion sensors and magnetic reed switch sensors. In [8], using temperature sensors only, they achieved 85% and 83% to predict two persons in a single room. Their accuracy can be up to 95% - 99% with all predictors including light, $CO_2$ and humidity sensors.

An estimation algorithm based on unsupervised clustering of both overlapped and non-overlapped conversational data with a change point detection algorithm for locomotive motion of the users to infer the occupancy was proposed as a mobile app in [30]. Users installed the app and provided consent for the app to access the smartphone sensors data. Using the random forest (RF) algorithm, they applied occupancy detection and the accuracy was 76% for counting a maximum number of 8 people.

A framework was created to produce occupancy estimates at different levels of granularity and provide confidence measures for effective building management in [29]. By using KNN and SVM, their accuracy varied from 64.6% up to 94.7%.

A real-time occupancy detection by using decision trees (DT) with multiple types of sensors such as light,

sound, $CO_2$, motion and computer power sensor was conducted by [25]. The lowest accuracy was received from sound sensors (90.79%) and the highest one from motion sensors (98.44%).

A model for real-time estimation of building occupancy sensing was presented by [16, 17]. Both papers utilised ANN and ran in the Waikato Environment for Knowledge Analysis (WEKA) and MATLAB for an open plan office occupancy detection and the accuracy is between 75% and 84.59%. With ambient sensors such as temperature, relative humidity, sound, light and $CO_2$, they used volatile organic compounds (VOCs) data as one of their features.

A systematic approach to occupancy modelling by using various ambient sensor data for both single occupancy and multi-occupancy room was proposed by [49]. With ANN, DT, KNN, naïve Bayes (NB), tree augmented naïve Bayes network (TAN) and SVM, human occupancy can be predicted with up to 98.2% in DT.

By utilising smart phone sensor such as microphone, Bluetooth and WiFi, one study [24] obtained a precision and recall of over 80% with 45 participants using mobile crowd sensing on group-aware recognition.

A model that utilises temperature, humidity, $CO_2$, sound, pressure and illumination sensor was used to detect whether the room is occupied or vacant in [3]. The occupancy detection was improved by using feature engineering and various machine learning algorithms such MLP, GP with RBF, SVM, RF and NB were compared. The human occupancy can be detected above 95% with multiple machine learning algorithms such as MLP, GP-RBF and RF.

Table 4: Algorithms, devices and reported accuracies for Carbon Dioxide sensor-based occupancy detection research.

| Source | Occupancy Detection Algorithms | Detailed Devices | Max. # of People | Accuracy (Occupancy) |
|---|---|---|---|---|
| [11] | Time-lagged mass balance approach | PP Systems SBA-5 $CO_2$ Gas Analyzers | 3 | N/A |
| [7] | Mass Balance Equation | $CO_2$ sensors | 12 | Binary prediction: 95.8%<br>Counting prediction: 80.6% |
| [5] | Non-negative matrix factorization (NMF)<br>Ensemble Least Square Regression<br>Support Vector Regression (SVR) | $CO_2$ sensors (BACNet server) | 15 (lab)<br>42 (classroom) | Binary prediction: 91%<br>Counting prediction: 15% |
| [25] | Decision Trees (DT) | $CO_2$ sensors | 1 | 94.68% |
| [32] | Hidden markov models<br>Neural networks<br>Support Vector Machines (SVM) Latent | Gas detection $CO_2$ sensor network | 4 | 65% - 80% |

## 2.6. $CO_2$-based Indoor Human Occupancy Detection

Since $CO_2$ sensors are already integrated with the BMS and ventilation infrastructure, we focus on utilising only $CO_2$ sensor data to estimate the number of indoor human occupancy. Consequently, the operational cost can be reduced by not purchasing and installing extra sensors such as PIR or motion sensors. Table 4 presents a summary of related work on indoor occupancy detection using $CO_2$ sensors.

Machine learning algorithms including HMM, NN and support vector machine latent (SVM latent) were implemented in [32] by using $CO_2$ data with the sensors deployed both inside and outside a room. By feature engineering $CO_2$ data with first order and second order difference of $CO_2$, the accuracy achieved is between 65% - 80% for binary occupancy prediction.

A method to assess human occupancy and occupant activity estimation in ten hospital rooms was conducted in [11] as part of the Hospital Microbiome Project. Using time-lagged mass balance approach, Dedesko el. al. aimed to determine occupant characteristics and understand the interactions between humans and microbial communities. The accuracy result was not mentioned in the paper.

CO2 based occupancy detection in office and residential buildings was conducted in [7]. The testing and validation have been done for both residential and non-residential buildings. Mass Balance Equation was implemented and vacant-occupied prediction accuracy is 95.8% and human counting prediction accuracy is 80.6%.

A heterogeneous sensor array was equipped in office workspace for the purpose of a real-time occupancy detector [25]. Decision Trees was implemented to perform the classification and to explore the relationship between different types of sensors, features derived from sensor data, and occupancy. The prediction accuracy for human occupancy using $CO_2$ sensor is 94.68%, lower than both motion sensor (98.44%) or current sensor (96.27%), but higher than light sensor (81.02%) and sound sensor (90.79%).

PerCCS is a model with a non-negative matrix factorization method to count people [5] using only one predictor in $CO_2$. In predicting vacant occupancy, they achieved up to 91% but only 15% accuracy in predicting the number of occupants.

Table 5: Example of Indoor Human Occupancy History Data

| Timestamp (t) | $CO_2$ (ppm) | Occupancy (person) | Activity |
|---|---|---|---|
| 18/05/2015 09:36:53 AM | 457 | 0 | (room empty) |
| 18/05/2015 10:01:55 AM | 550 | 1 | (arrived) |
| 18/05/2015 10:41:59 AM | 596 | 0 | |
| 18/05/2015 11:37:05 AM | 580 | 3 | (group meeting) |
| 18/05/2015 12:07:07 PM | 725 | 3 | |
| 18/05/2015 12:32:10 PM | 500 | 0 | (lunch break) |
| 18/05/2015 12:42:11 PM | 510 | 0 | |
| 18/05/2015 12:47:12 PM | 514 | 1 | (back from lunch) |
| 18/05/2015 02:02:23 PM | 508 | 0 | (went to seminar) |
| 18/05/2015 03:52:35 PM | 503 | 1 | |
| 18/05/2015 03:02:30 PM | 397 | 0 | (external meeting) |
| 18/05/2015 04:26:55 PM | 570 | 1 | |
| 18/05/2015 05:38:22 PM | 475 | 0 | (went home) |
| ... | ... | ... | |

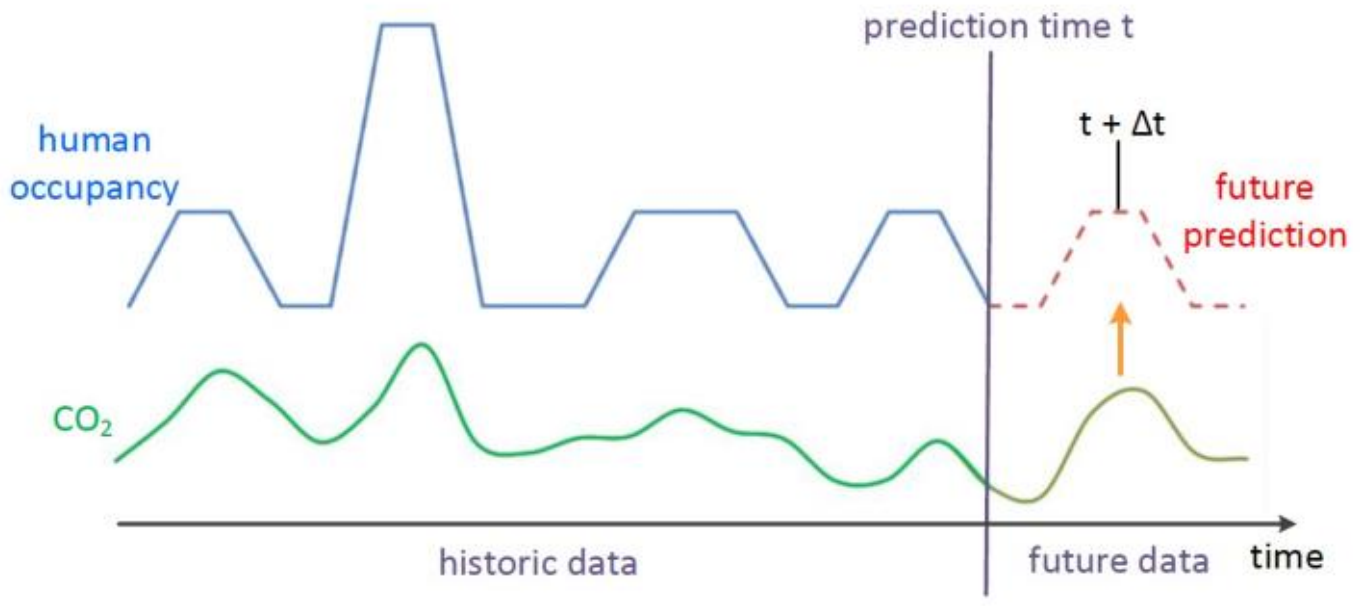

Figure 3: Real-time prediction scenario for continuous t showing the amount of $CO_2$ fluctuations. The fundamental task is to predict the number of occupants at time $t+\Delta t$.

### 2.7. *Summary of Related Work*

After reviewing all the literature in occupancy detection, a few key points are derived as follows:

- PIR is the standard sensor for indoor human occupancy. PIR sensors are cheap, but they are not scalable. To be able to predict occupancy for 100 rooms, at least 100 PIR sensors need to be provided for the purpose of counting the people;
- not every research paper gives their accuracy. This makes the direct comparison between each model difficult;
- poor sensor calibration and lower frequency reduce the accuracy prediction result;
- $CO_2$ is one of the best types of sensors to measure indoor human occupancy, but the majority of indoor human occupancy papers with $CO_2$ are binary occupancy analysis, not real people counting analysis;

- using too many types of sensors as features can result in lower accuracy;
- there is no single formula available to calculate the number of occupants, and from subsection 2.1 we conclude that collecting all the parameters related to indoor human occupancy is not practical and is subject to privacy breaches;
- the majority of previous experiment datasets are not publicly available;
- ambient sensor research is done without the involvement of user consent (the cons of smartphone-based apps).

Overall, sensor-based detections have higher accuracy compared to radio based detections. For example, Wi-Fi and RSSI signals achieved 63% accuracy for indoor detection [12] with 9 occupants. For occupancy counting, $CO_2$ sensors only have been experimented with the maximum of 42 occupants and accuracy limit of 15%[5].

## 3. Problem Definition

In Table 5 and Figure 3, the data shows there is a dependency between $CO_2$ and occupancy data. Our research question is how can we predict the number of people by using a single $CO_2$ sensor with the accuracy similar to the state-of-the-art techniques in the occupancy detection field?

### 3.1. Scenario Assumption

Assume TS represents the length of a time series and is expressed as $TS = \{ts_1, ts_2, ..., ts_q\}$, where q means the number of sample points. In our time series datasets, we have two aspects:

- Carbon dioxide ($CO_2$) concentration $C$, defined as
  $C = \{C_1, C_2, ..., C_q\}$
- Indoor human occupancy $O$, the number of people in the room at TS defined as
  $O = \{O_1, O_2, ..., O_q\}$

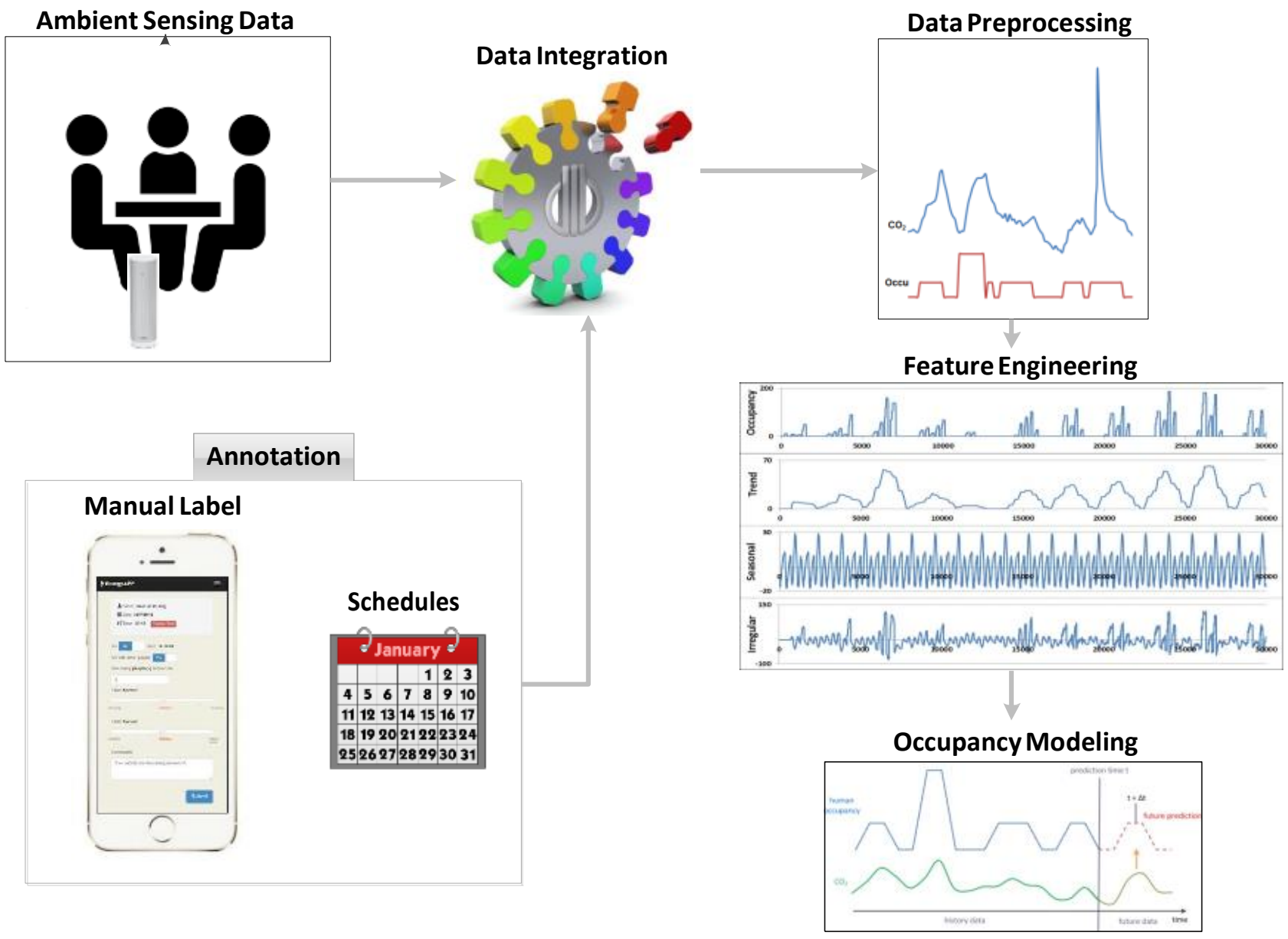

Figure 4: Data Collection and Analysis Framework

### *3.2. Problem Definition*

In time series prediction, analysing one-step-ahead prediction is different from analysing multi-step-ahead prediction. Predicting multi-step-ahead needs a more complicated method due to the accumulation of errors and the number of uncertainties increasing with time. We focus on multi-step-ahead prediction with the support of one dependent variable to reduce uncertainties.

We have two different types of datasets: $CO_2$ concentration $C$ and indoor human occupancy $O$. In order to explore the relationship between both factors above, there are two problems that need to be solved:

- Decompose both $CO_2$ concentration and indoor human occupancy to reduce the level of complexity.
- Explore the correlation between $CO_2$ concentration and indoor human occupancy and all of their decomposed components to find what correlations exist between each component.

## 4. Method

We assess our model in two different locations with very different contexts to ensure this model works under various conditions. The first location is one academic office belonging to one staff member at RMIT University, Australia. This room is chosen for human occupancy prediction since a controlled experiment can be conducted for an extended period of data collection. The picture of the room is shown in Figure 5. The second dataset was collected inside a cinema theatre in Mainz, Germany [47]. Cinema theatre is chosen as another setting due to its nature of having fluctuating numbers of people throughout the day. The numbers of people in the audiences can reach hundreds and can decrease to zero within a few hours.

### *4.1. Data Collection Experiment Setup*

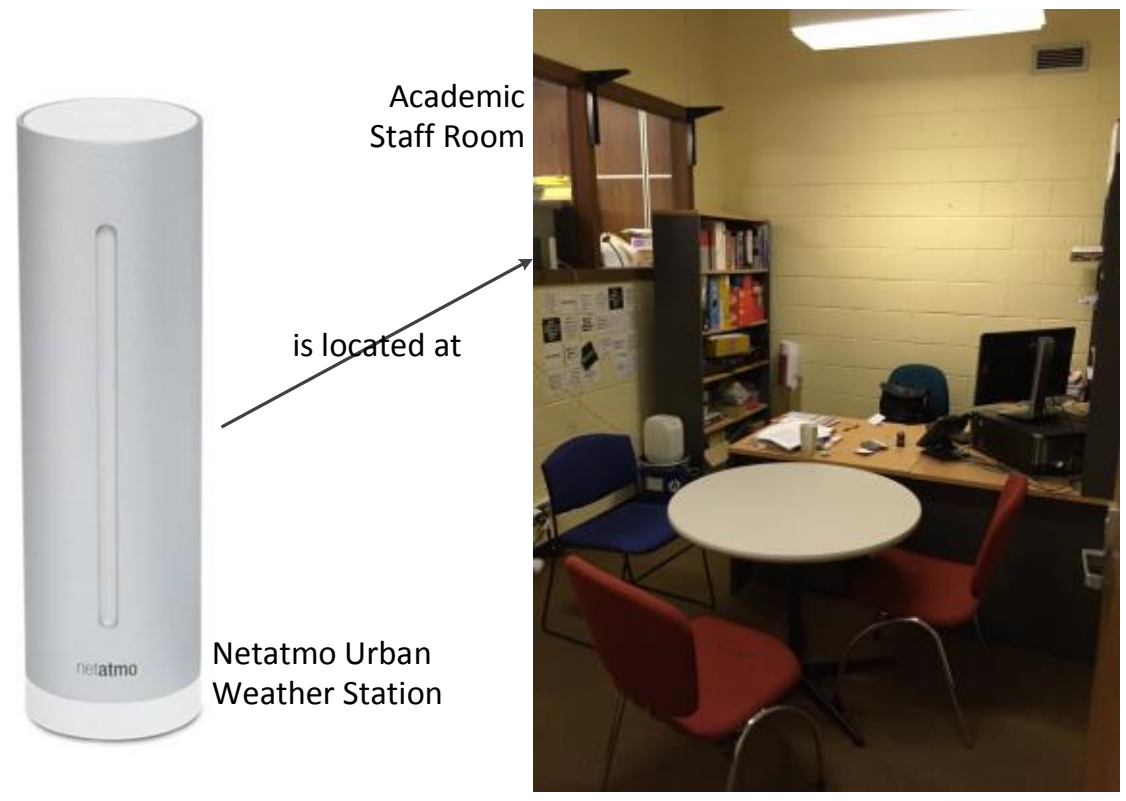

Figure 5: The left picture is a Netatmo urban weather station, a sensor device to gather ambient $CO_2$ data which was set up near the window in the academic staff room (right).

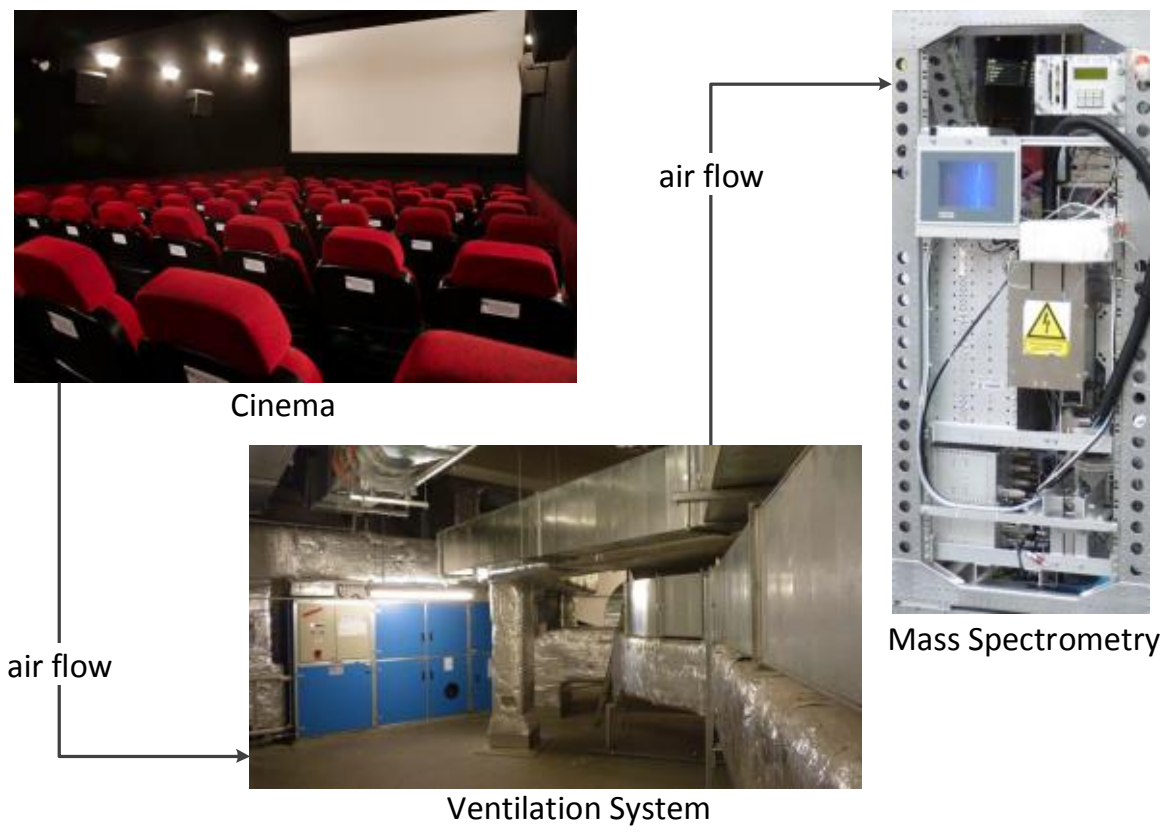

Figure 6: Measurement in the cinema theatre. Air is drawn out through the screen room via the ventilation system and is transported to the mass spectrometer. [47]

#### *4.1.1. Academic Staff Room*

We use a commercial off-the-shelf Netatmo urban weather station (Range: $0 - 5000$ ppm, accuracy: $\pm 50$ ppm) to read and collect ambient $CO_2$ data, shown in Figure 5. The experiment took place between May and June 2015 and the data frequency is 5 minutes. The dataset is uploaded to a cloud service for integration purposes. Due to the small room characteristic (3x4x5m), N for time lag is 0 as we assume that there is a negligible time period between exhaling process and sensor reading. We selected 2 weeks data from the whole dataset and used them in the further analysis. To obtain actual occupancy data for this room, one staff volunteered to manually label the occupancy for the whole duration of the experiment.

#### 4.1.2. Cinema Theatre

The cinema dataset were collected between December 2013 and January 2014 [47]. The air-flow measurement and device arrangement for the cinema dataset is shown at Figure 6. The dataset was collected using mass spectrometry machinery installed on the air ventilation system. The air flows from the screening room via the ventilation system to the mass spectrometer (Range: $10 - 500$ m/z with a Time-of-flight (TOF) acquisition sampling time per channel of 0.1 ns, resolution: $\pm 3700$ m/$\Delta$m) for data analysis. To obtain actual occupancy data for this cinema theatre, we use the number of tickets that were sold for each movie session.

#### 4.1.3. Experiment Tool

We utilised WEKA, MATLAB and R to help us perform this experiment. WEKA is used for polynomial linear regression with the M5 method for both correlation models for trend (Subsection 4.3.2.1) and irregular features (Subsection 4.3.2.3). MATLAB code is run for the baseline method, SVR and its prediction result. We used R to integrate all the data, including decomposition of STD and STL and the majority of data pre-processing. We imported the data from R into Microsoft Excel for data analysis and visual output.

### 4.2. Data Pre-processing

This section explains our data pre-processing and the reason why data pre-processing is important for our model. We gathered both the $CO_2$ concentration from the sensor data and indoor human occupancy from our annotations as shown in Figure 4. Both data are integrated and pre-processed using our pre-processing method described below in Subsection 4.2.1 (autocorrelation and the line of best fit) and Subsection 4.2.2 (time lag). With feature engineering, we factorised the data into different features and applied prediction models described in Section 4.3 to predict the indoor human occupancy.

To solve a time delay issue between $CO_2$ data and our prediction indoor human occupancy number, the data needed to be pre-processed. The time delay issue means that when one person enters a room, it will take some time before the $CO_2$ level in the air increases proportionally.

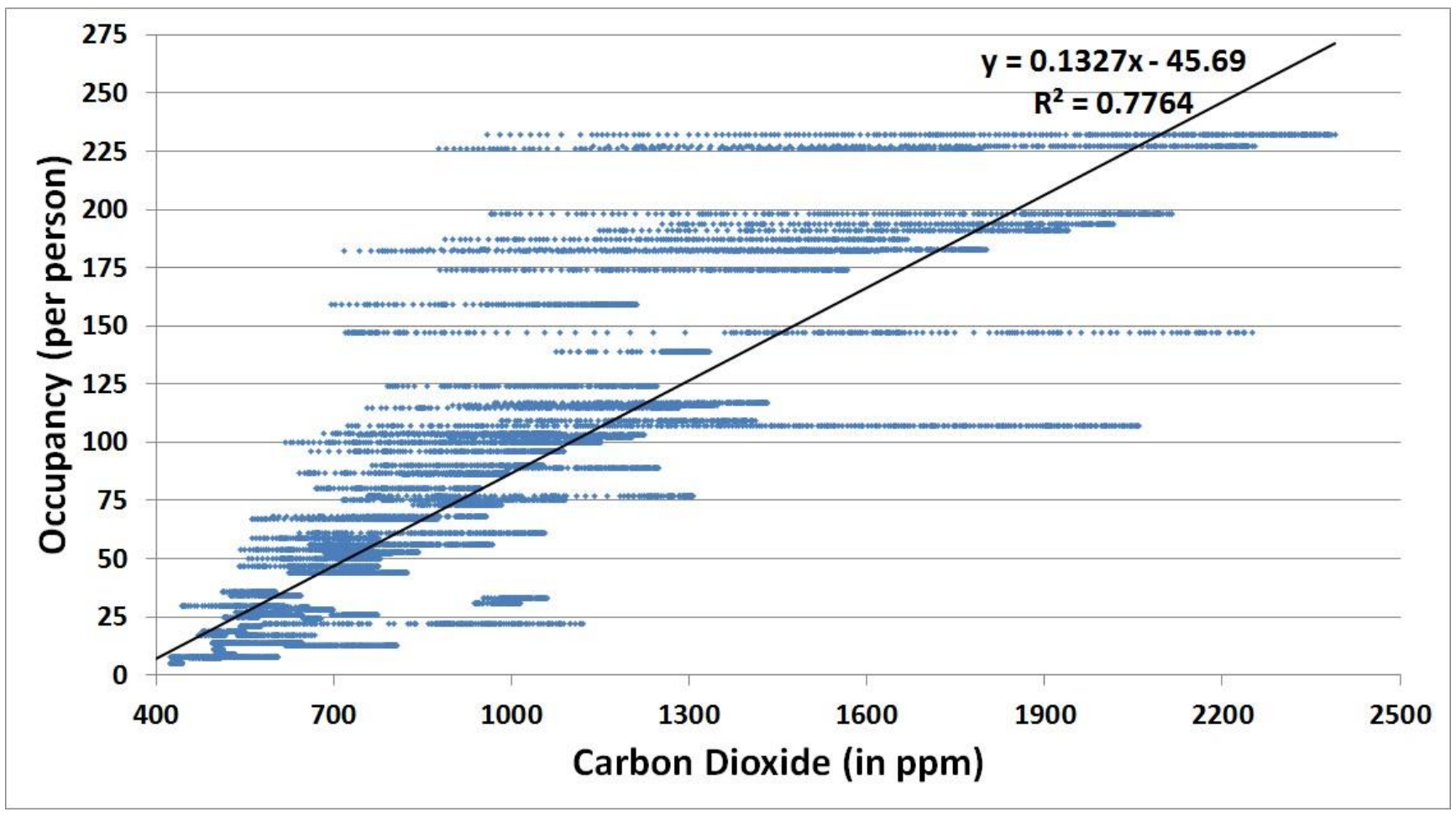

Figure 7: Correlation between the number of occupants and $CO_2$ readings in ppm with time lag 0.

#### 4.2.1. Autocorrelation and the Line of Best Fit

In order to analyse the data between $CO_2$ and the number of occupants, a data lagging issue needs to be considered. Data lagging means that it will take some time for $CO_2$ to populate the room as there is a delay between the time of people entering (or exiting) the room and the increment (or decrement) of the $CO_2$ reading. For each dataset from 0 minutes time lag to Upper Bound (UB) minutes time lag, the correlation of $CO_2$ data with the number of occupancies is drawn. UB value is defined by the formula in Equation 2.

$$UB = ||(roomlength * roomwidth * roomheight)/100|| \tag{2}$$

For the academic staff room, the UB value is 1. This small value means the usefulness of time lag for this room analysis is minimal. For cinema data [47], the UB value is 60 due to the large size of the studio in the cinema theatre. A line of best fit is drawn as shown in Figure 7.

To calculate a line of best fit, first we calculate the slope value between $CO_2$ and occupancy data. The formula is defined in Equation 3.

$$SL = \frac{\sum (O_t - \bar{O_t})(C_t - \bar{C_t})}{\sum (O_t - \bar{O_t})^2} \tag{3}$$

| | |
|---|---|
| $SL$ | slope of the linear regression line |
| $O_t$ | occupancy value |
| $\bar{O_t}$ | sample means (the averages) of the known occupancy value |
| $C_t$ | $CO_2$ value |
| $\bar{C_t}$ | sample means (the averages) of the known $CO_2$ value |

After getting the SL value, intercept value between both data needs to be calculated with formula in Equation 4.

$$INTC = \bar{C_t} - SL * \bar{O_t} \tag{4}$$

| | |
|---|---|
| $INTC$ | intercept (the value at the intersection of the y axis) of the linear regression line |
| $\bar{C_t}$ | sample means of the known $CO_2$ value |
| $\bar{O_t}$ | sample means of the known occupancy value |

The main formula for the line of best fit (LBF) is shown in Equation 5.

$$LBF = (O_t - (SL * C_t + INTC))^2 \tag{5}$$

| | |
|---|---|
| $O_t$ | occupancy value |
| $SL$ | slope of the linear regression line |
| $C_t$ | $CO_2$ value |
| $INTC$ | intercept of the linear regression line |

#### *4.2.2. Time Lag*

For each line of best fit, the mean squared error (MSE), root-mean-square deviation (RMSD) and the normalised root mean squared error (NRMSE) are calculated. The formula for calculating NRMSE is shown in Equation 6.

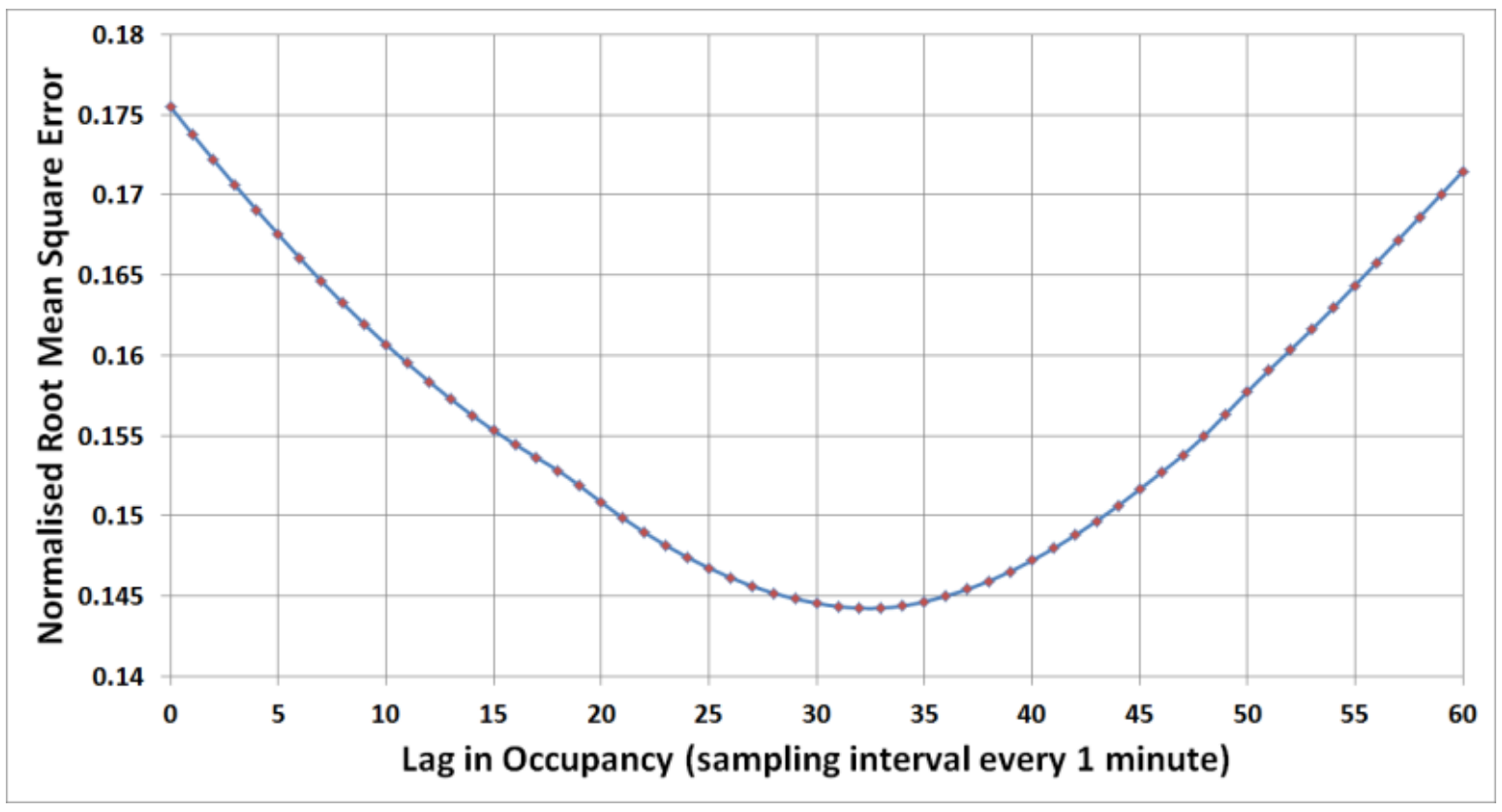

Figure 8: Ordinary Least Square Regression Normalised Root Mean Squared Error (NRMSE) between $CO_2$ data and actual occupancy for 60 minutes time lag.

$$NRMSE = \frac{\sqrt{\frac{1}{n}\sum_{t=1}^{n}(C_t - \bar{C}_i)^2}}{O_{max} - O_{min}} \quad (6)$$

| | |
|---|---|
| $NRMSE$ | normalized root mean square error |
| $t$ | total number of data set |
| $C_t$ | $CO_2$ value |
| $\bar{C}_t$ | sample means of the known $CO_2$ value |
| $O_{max}$ | maximum occupancy value |
| $O_{min}$ | minimum occupancy value |

This step is repeated at UB times for each time lag. For time lag analysis, we use least square regression to compare each NRMSE from time lag 0 until time lag UB. We use the lowest number of NRMSE value, time lag value (TL), as our baseline time lag for the data analysis shown in Equation 7.

$$TL = min(NRMSE) \quad (7)$$

For the academic staff room, the TL value is 0. This value represents no time lag is needed for this analysis. For the cinema theatre, the lowest number of error value happens at time lag TL=32 as shown in Figure 8. This TL value is our base for the cinema theatre data analysis. So for the entire cinema data analysis process, we use time lag 32.

### *4.3. Human Occupancy Counting Algorithm*

Since there is no linear relationship between $CO_2$ and indoor human occupancy, we introduce a new prediction model to address this non-linear correlation issue by decomposing both $CO_2$ and occupancy data. There are two variants of this model, CD-HOC seasonal trend decomposition (CDHOC-STD) and CD-HOC seasonal trend decomposition based on Loess (CDHOC-STL), which will be discussed in the following subsections. Each decomposition extracts a feature and correlation can be developed from each feature.

The next following subsections explain the core prediction model for this paper, shown in Figure 9. In the first subsection, we discuss two data decomposition methodologies. The next subsection explains the correlation model for trend, seasonal and irregular features. The last subsection presents a new method for analysing conditions when the room is vacant which we term zero pattern adjustment, to increase the overall accuracy. This model needs to be re-trained for each location to obtain the best accuracy result.

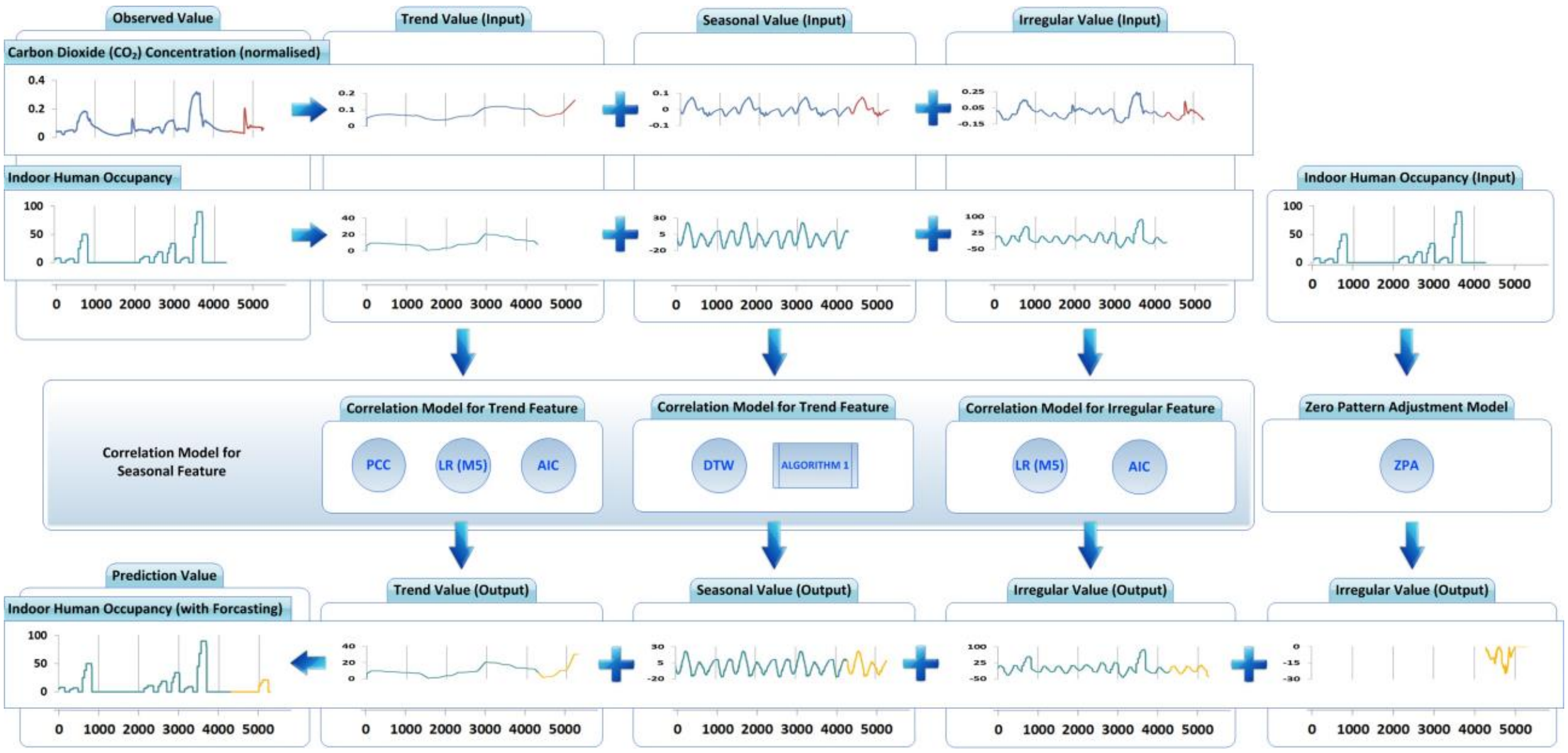

Figure 9: Indoor Human Occupancy Prediction Model, CD-HOC

#### *4.3.1. Decomposition Methodologies*

There are two variants of decomposition methods that are utilised for CD-HOC, seasonal-trend decomposition (STD) in subsection 4.3.1.1 and seasonal-trend decomposition based on Loess (STL) in subsection 4.3.1.2.

*4.3.1.1. Seasonal-Trend Decomposition.* STD is a mature technique in time series analysis. One of the most popular variants is the version X-11 method for using moving average [43] and the most recent variant is the version X12-ARIMA [19]. STD is an integral part of our framework.

To understand the time series data better, we use STD to decompose the model into four main features: trend, cyclical, seasonal and irregular. The trend feature ($T_t$) reflects the long-term progression of the time series during its secular variation. The cyclical feature ($C_t$) is a repeated but non-periodic fluctuation during an extended period of time. The seasonal feature ($S_t$) is a systematic and regularly repeated sequence during a short period of time. The irregular feature ($e_t$ also known as error or residual) is a short-term fluctuation from the time series and is the remains after the trend, cyclical and season features have been removed. For this paper, as our experiment is within a short period of time (one month for each case), we decide to combine the cyclical feature into trend feature to make the model simpler without sacrificing the accuracy.

Below is the core logic for STD:

1. Calculate 2x12 moving average in the raw data (both $CO_2$ and occupancy datasets) to obtain a rough trend feature data $T_t$ for all period.
2. Calculate ratios of the data to trend, named "centred ratios" ($y_t/T_t$).
3. To form a rough seasonal feature ($S_t$) data estimation, apply separate 2x2 moving average to each month of the centred ratios.
4. To obtain the irregular feature ($e_t$), divide the centred ratios by $S_t$.
5. Multiply modified $e_t$ by $S_t$ to get modified centred ratios.
6. Repeat step 3 to obtain revised $S_t$.
7. Divide the raw data by the new estimate of $S_t$ to give the preliminary seasonal adjusted series, $y_t/S_t$.
8. The trend feature ($T_t$) is estimated by applying a weighted Henderson moving average to the preliminary seasonally adjusted values.
9. Repeat step 2 to get new ratios by dividing the raw data by the new estimate of $T_t$.

10. Repeat Steps 3-5 using the new ratios and applying a 3x5 moving average instead of a 3x3 moving average.
11. Repeat step 6 but using 3x5 moving average instead of a 3x3 moving average.
12. Repeat step 7.
13. Finally the reminder feature is obtained by dividing the seasonally adjusted data from step 12 by the trend feature obtained in step 8.

Our customised STD formulation is:

$$STD_t = f(T_t, S_t, e_t) \tag{8}$$

| | |
|---|---|
| $t$ | time |
| $STD_t$ | actual value of a time series at time t |
| $T_t$ | trend feature at t |
| $S_t$ | seasonal feature at t |
| $e_t$ | irregular feature at t |

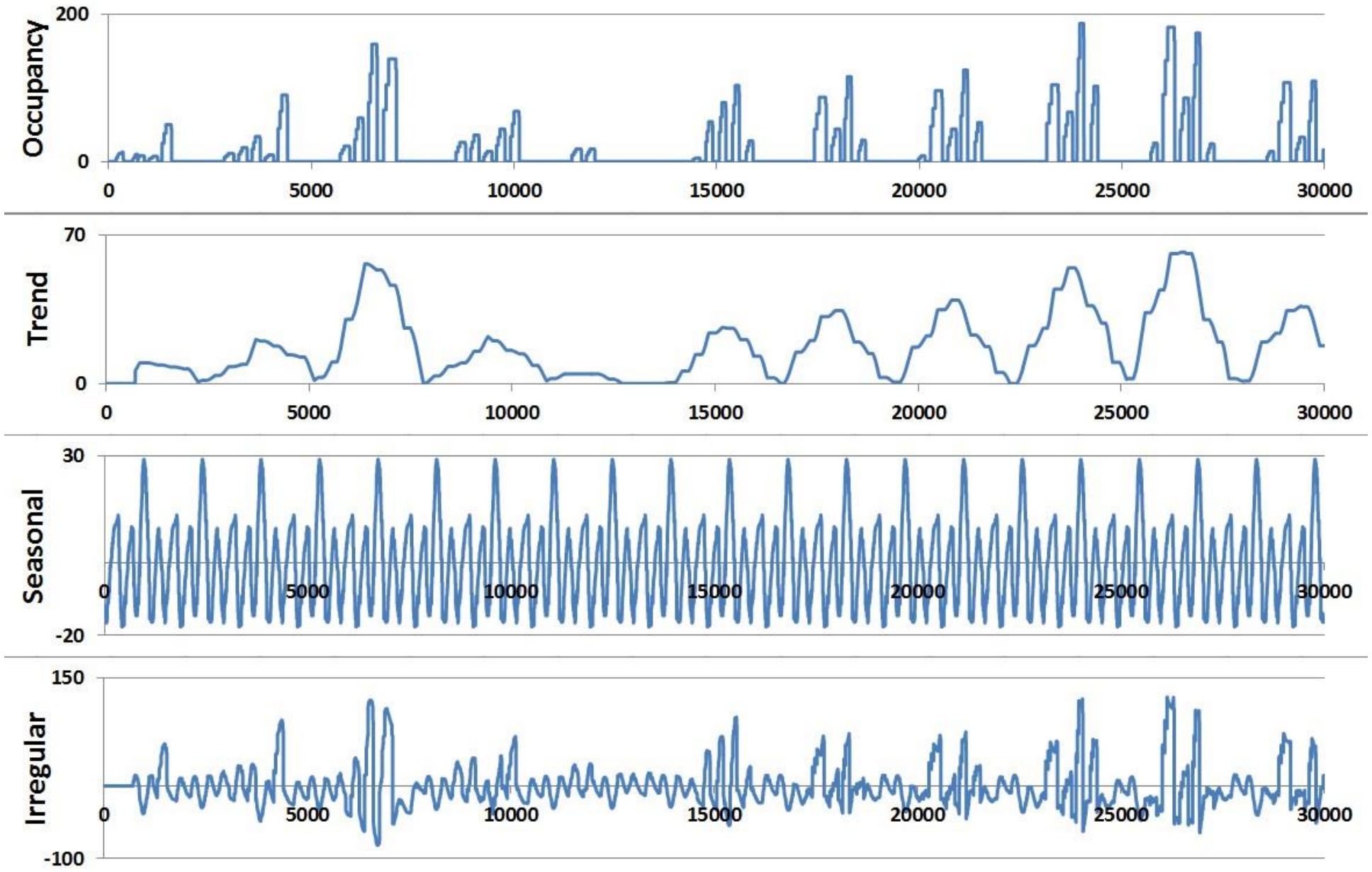

Figure 10: Example of Time Series Decomposition

The function $f()$ can be additive or multiplicative, yielding an additive decomposition or a multiplicative decomposition. Additive decomposition model is a data model in which each factor is added to model the data. Multiplicative decomposition occurs when the seasonal feature pattern is increased as the number of data increases. For multiplicative decomposition, the trend and seasonal features are multiplied and then added to the irregular feature. In our case, as the magnitude of seasonal feature pattern in the data does not depend on the magnitude of the overall dataset, we decided to use additive decomposition as shown in Figure 10. From Equation 8, our general STD formula becomes:

$$STD_t = T_t + S_t + e_t \tag{9}$$

This general STD formula will be applied to both time series for the $CO_2$ dataset and the human occupancy dataset:

$$C_t = T_t^C + S_t^C + e_t^C \tag{10}$$

$$O_t = T_t^O + S_t^O + e_t^O \tag{11}$$

To predict $O_{t+1}$ up to $O_{t+n}$, we need to create a model to systematically predict each of $T_{t+1}^O$, $S_{t+1}^O$ and $e_{t+1}^O$ up to $T_{t+n}^O$, $S_{t+n}^O$ and $e_{t+n}^O$ and then reconstruct the new prediction dataset using the additive method. In this paper, we explore two variants of STD for our model and compare the accuracy result with the baseline. The first one is standard STD that is implemented using moving average and the second one is seasonal-trend decomposition based on Loess (STL) [10].

*4.3.1.2. Seasonal-Trend Decomposition based on Loess.* Loess stands for locally weighted scatterplot smoothing, a non-parametric local regression method that is built on least square regression. This method is designed to estimate a non-linear relationship in a data set. STL is a filtering procedure for decomposing a time series into trend, seasonal and irregular components with loess smoother. To be compared with other STD variants such as X-12 ARIMA, STL has several advantages namely:

- STL method is very versatile and robust. It can handle any type of seasonality and will not be limited to a monthly or quarterly dataset;
- The seasonal feature and the smoothness of the trend-cycle both can be controlled by the user;
- STL is robust on outliers, so occasional unusual data will not ruin the estimation of trend-cycle and seasonal features. However, they will affect the irregular feature.

### 4.3.2. Correlation Models

There are three correlation models below for trend features in subsection 4.3.2.1, seasonal features in subsection 4.3.2.2 and irregular features in subsection 4.3.2.3.

*4.3.2.1. Correlation Model for Trend Feature ($T_t$).* As the definition of the trend feature ($T_t$) is the long-term non-periodic progression of the time series during its secular variation, we assume that the trend feature for the $CO_2$ dataset ($T_t^C$) will be similar to the trend feature for indoor human occupancy ($T_t^O$).

To check the similarity between both trend features, we use the Pearson product-moment Correlation Coefficient (PCC) as shown below:

$$r = \frac{n(\sum xy) - (\sum x)(\sum y)}{\sqrt{[n\sum x^2 - (\sum x)^2][n\sum y^2 - (\sum y)^2]}} \tag{12}$$

$r$ correlation coefficient
$x$ dataset x
$y$ dataset y
$n$ number of sample points

The range of Pearson's r value is from -1 to +1. If the value is >0.7, the correlation between both datasets is strongly positive.

Once the validation step is done, we implement polynomial M5 linear regression. The M5 method will build trees whose leaves are associated with multivariate linear models and the nodes of the tree are chosen over attributes that maximise the expected error reduction, given by the Akaike Information Criterion (AIC - a measure to check the relative goodness of fit of a statistical model) [2]. The purpose of using AIC is to evaluate the model. The value for each of trend feature needs to be a positive value so we put the absolute value on both the $CO_2$ ($|T^C_t|$) and human occupancy trend features ($|T^O_t|$). The main formula for trend feature correlation is shown below:

$$T^O_t = \alpha_0 + \alpha_1(T^C_t) + \alpha_2(T^C_t)^2 + ... + \alpha_n(T^C_t)^n + E \tag{13}$$

Linear regression with M5 will output each $\alpha_n$ and $E$ value. With these parameters, the future trend for $T^O_{t+n}$ can be obtained.

**Algorithm 1** Finding a repeated pattern sequence inside seasonal feature

**procedure** REPEATED_SEQUENCE($S_t$)<br>
    $s^{temp}_t, s^{fin}_t \subset S_t$<br>
    $len \leftarrow 0$ ▷ $len$: Length for $s^{temp}_t$<br>
    $a \leftarrow S_t[len]$ ▷ $a$: Start Point<br>
    **for** each node $i \in S_t$ **do**<br>
        $len{+}{+}$<br>
        $s^{temp}_t \leftarrow s^{temp}_t + S_t[i]$<br>
        **if** $a = S_t[i]$ **then**<br>
            **if** DTW($s^{temp}_t$, $S_t[i+1..i+len]$) > 95 **then**<br>
                $s^{fin}_t \leftarrow s^{temp}_t$<br>
                **break**<br>
            **end if**<br>
        **end if**<br>
    **end for**<br>
    **return** $s^{fin}_t$<br>
**end procedure**

*4.3.2.2. Correlation Model for Seasonal Feature ($S_t$).* The seasonal feature ($S_t$) is a systematic and regularly repeated sequence during a short period of time. Due to this characteristic, every seasonal feature can be fitted by a finite Fourier series. To correlate $S^C_t$ and $S^O_t$, we use Dynamic Time Warping (DTW), a pattern matching technique to score the similarity between the shape of particular signal within certain duration [40]. The full correlation algorithm is implemented in Algorithm 1 to find regularly repeated sequences within each $S_t$.

Once we find a sequence that is repeating in $s^{fin}_t$ for both the $CO_2$ and occupancy seasonal features, we compare the length of $s^{fin(O)}_t$ and $s^{fin(C)}_t$. If the length of $s^{fin(O)}_t < s^{fin(C)}_t$, we apply an interpolation method inside $s^{fin(O)}_t$, so both have the same length. If the length of $s^{fin(O)}_t > s^{fin(C)}_t$, we apply data reduction method so finally both have the same length. The final regression equation for seasonal feature correlation is shown below:

$$s^{fin(O)}_t = f(s^{fin(C)}_t) \tag{14}$$

With this equation, the future trend for $S^{fin(O)}_{t+n}$ can be obtained.

*4.3.2.3. Correlation Model for Irregular Feature ($e_t$).* Due to similar characteristics between trend and irregular features, we apply the same correlation method from the trend feature:

$$e^O_t = \beta_0 + \beta_1(e^C_t) + \beta_2(e^C_t)^2 + ... + \beta_n(e^C_t)^n + \gamma \tag{15}$$

The only difference from the trend feature is that we do not need to validate it using PCC as the shape of the irregular feature will depend more on its trend and seasonal features.

*4.3.2.4. Zero Pattern Adjustment.* In human occupancy prediction research, inferring knowledge when a room is vacant is paramount. By minimising false positives, the accuracy prediction can be improved. The Zero pattern adjustment (ZPA) method learns the behaviour from previous historical data and makes some smart adjustments for a vacant room when the normal algorithm returns incorrect prediction. The ZPA technique overlays all previous dataset and puts them on a single 24-hour x-axis chart to determine the earliest start and end points when the room is vacant each day during the night to dawn period. We symbolise ZPA as $zpa_t^O$.

### 4.3.3. Occupancy Model

Algorithm 2 contains the main occupancy model, where we integrated each feature to get the occupancy prediction value.

---

**Algorithm 2** Finding the indoor human occupancy

---

**procedure** NUMBER OCCUPANCY($T_t^O$, $S_t^O$, $e_t^O$, $zpa_t^O$)
    $O_t^{temp} \leftarrow T_t^O + S_t^O + e_t^O + zpa_t^O$    ▷ $O_t^{temp}$: temporary occupancy
    **if** $O_t^{temp} \geq 0$ **then**
        $O_t = O_t^{temp}$
    **else**
        $O_t = 0$
    **end if**
    return $O_t$
**end procedure**

---

## 5. Experiments, Results and Discussion

### 5.1. Experiment

There are two types of feature decomposition: CDHOC-STD and CDHOC-STL. We implement both CD-HOC models for each different combination of training and test dataset. Finally, we compare each CD-HOC accuracy result with SVR.

CD-HOC model predicts each future value for the whole period of time based on specific time window. To understand this model better and how well it performs compared with the baseline, we define $x$, accuracy error tolerance parameter. Zero units error tolerance means only the exact number recognised is considered as true positive. For example with ten units error tolerance, if the real indoor human occupancy is 150 people, the prediction shows that 146 or 155 is considered correct as it is within ±10 units error tolerance. The parameter $x$ value will be different based on the size of the room.

#### 5.1.1. Experiment parameters for academic staff room dataset

For the academic staff room dataset, we used 5-min time window. Total data that we gathered from this room are 4019 data spread in 14 days. Due to the small room size, we decided not to use time lag for data analysis. For this room, we have seven pairs of training-test datasets. It starts with seven days of training dataset and seven days of test dataset. It ends with 13 days of training dataset to predict one day test dataset.

#### 5.1.2. Experiment parameters for cinema dataset

For the cinema theatre dataset, we use a 3-minutes time window for data analysis. The data that we gathered from this cinema theatre consisting of 68640 instances spread over 23 days. The cinema theatre capacity is up to 300 people and for this experiment, we run the line of best fit for time lag 0 to time lag 60. The lowest normal root mean square error point is at time lag 32 and we use time lag 32 as time lag baseline. This time lag is appropriate as a bigger room needs a larger time lag for the model to have a better accuracy. For this room, we decided to use December 2013 data for training and January 2014 data for testing. Then we replicated it in the similar method by giving one day from testing dataset to training dataset and ran the model again. This method is repeated until the test dataset only consisted of one day of data. Finally, we ran the same training-test dataset using baseline method, SVR.

#### *5.1.3. Evaluation*

To evaluate the result, we divide the data into 2 equal parts. The first part is the training dataset and the second one is the test dataset. To be able to understand how well the model fits for a longer duration, we repeat the division of training and test dataset by adding one day data from the test dataset to the training dataset. This replication is repeated again until the test dataset has only one single day and the rest belong to training dataset. This incremental days of training and reduction in testing evaluation method ensures the robustness of the model. The baseline is state-of-the-art machine learning algorithm, SVR.

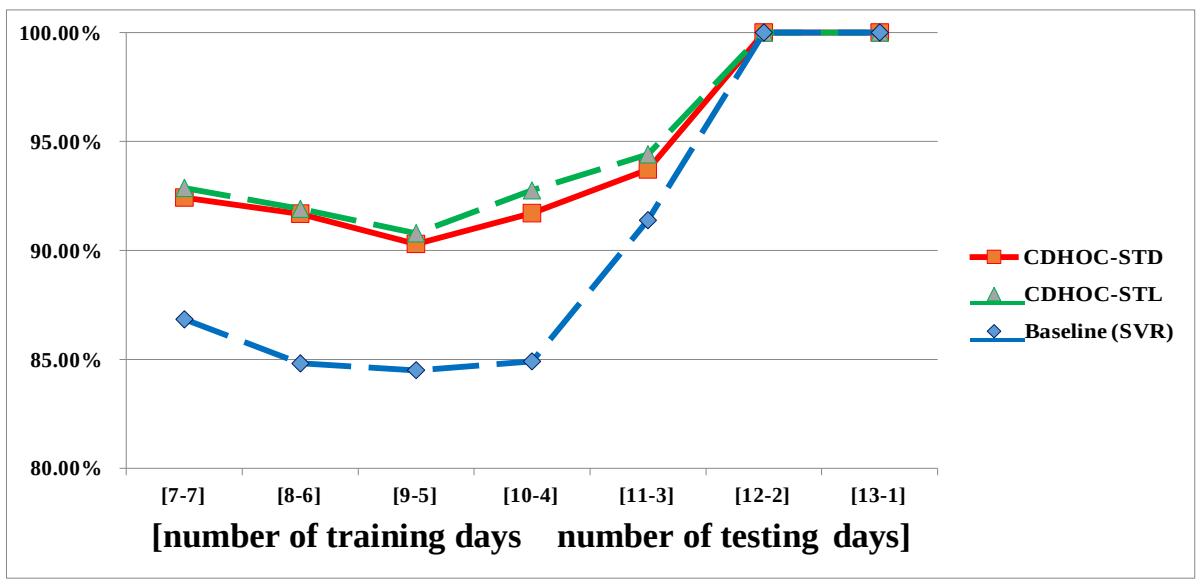

Figure 11: Academic staff room dataset - the comparison for indoor human occupancy with zero units error tolerance

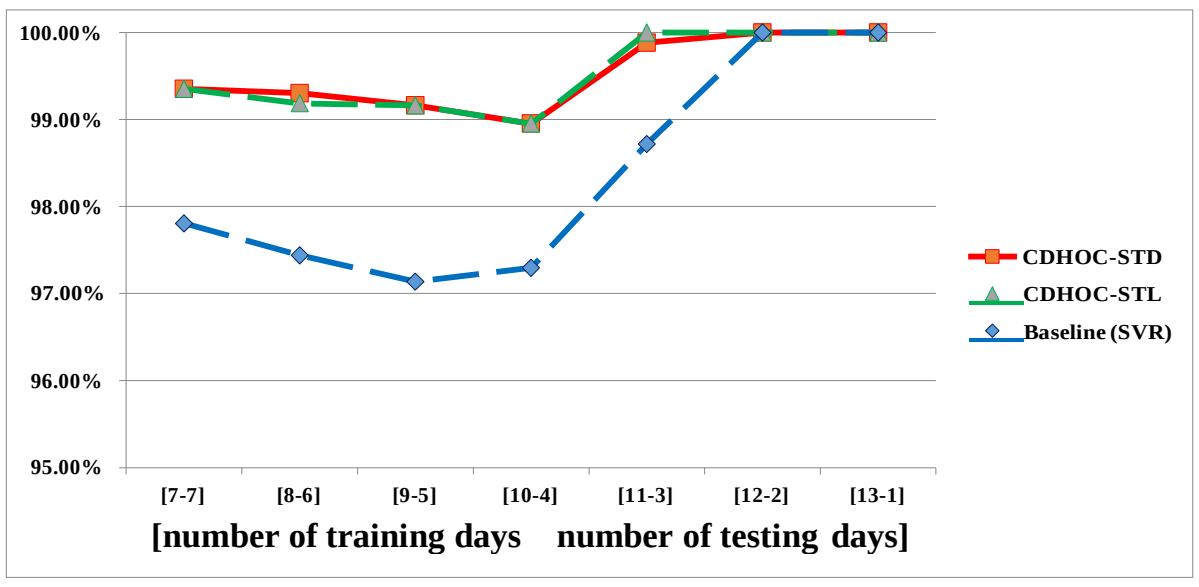

Figure 12: Academic staff room dataset - the comparison for indoor human occupancy with one units error tolerance

Table 6: Academic staff room indoor human accuracy result

| # of training days | # testing days | zero units error tolerance | | | one units error tolerance | | |
|---|---|---|---|---|---|---|---|
| | | Baseline (SVR) | CDHOC-STD | CDHOC-STL | Baseline (SVR) | CDHOC-STD | CDHOC-STL |
| 7 | 7 | 86.84% | 92.42% | 92.87% | 97.81% | 99.35% | 99.35% |
| 8 | 6 | 84.82% | 91.68% | 91.91% | 97.44% | 99.30% | 99.19% |
| 9 | 5 | 84.50% | 90.29% | 90.78% | 97.14% | 99.16% | 99.16% |
| 10 | 4 | 84.90% | 91.71% | 92.76% | 97.29% | 98.95% | 98.95% |
| 11 | 3 | 91.39% | 93.71% | 94.41% | 98.72% | 99.88% | 100.00% |
| 12 | 2 | 100.00% | 100.00% | 100.00% | 100.00% | 100.00% | 100.00% |
| 13 | 1 | 100.00% | 100.00% | 100.00% | 100.00% | 100.00% | 100.00% |
| Average accuracy | | 90.35% | 94.26% | 94.68% | 98.34% | 99.52% | 99.52% |

### *5.2. Experiment Result*

#### *5.2.1. Experiment Result for academic staff room dataset*

The average accuracy for indoor human occupancy with CDHOC-STD is 94.26%, with CDHOC-STL is 94.68% and with SVR is 90.35%. From Figure 11, both CDHOC-STD and CDHOC-STL performed better than the baseline on average by 4.33%. The last two days are Saturday and Sunday, so both CD-HOC and the baseline model correctly predict zero occupancies for each day.

To understand the model, we run the experiment and check the level of accuracy with one unit error tolerance. The accuracy result is shown in Figure 12. For this experiment, the maximum number of people is four at one time. The average accuracy with one unit error tolerance for indoor human occupancy with CDHOC-STD is 99.52%, with CDHOC-STL is 99.52% and with SVR is 98.34%.

#### *5.2.2. Experiment Result for cinema dataset*

For the cinema dataset, the comparison accuracy result is shown in Figure 13. Both CD-HOC methods perform better than the baseline method and on average each CD-HOC method has 8.5% higher accuracy in predicting indoor human occupancy. The highest prediction accuracy was found when we used 22 days data for training to predict the number of human occupants the next day.

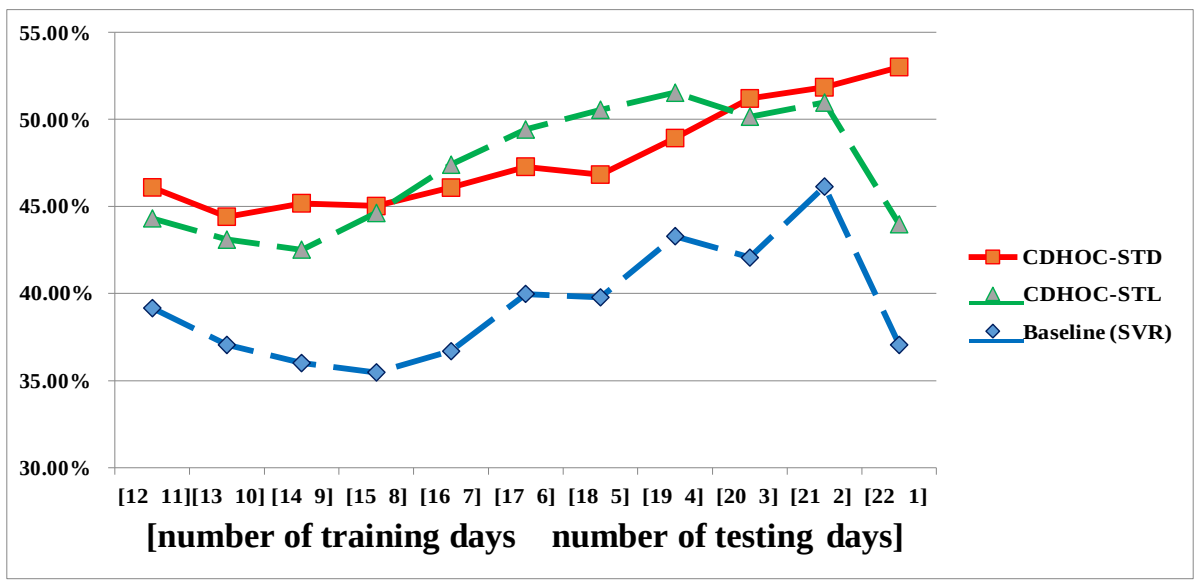

Figure 13: Cinema dataset - the comparison for indoor human occupancy with zero units error tolerance

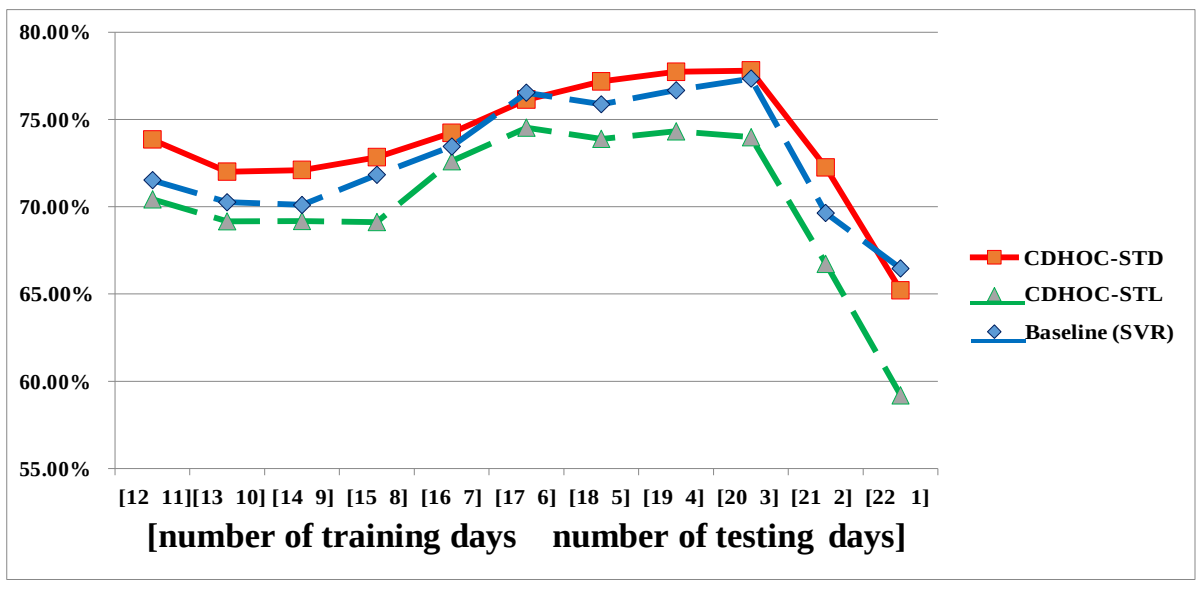

Figure 14: Cinema dataset - the comparison for indoor human occupancy with ten units error tolerance

Table 7: Cinema theatre indoor human accuracy result

| # of training days | # testing days | zero units error tolerance | | | one units error tolerance | | |
|---|---|---|---|---|---|---|---|
| | | Baseline (SVR) | CDHOC-STD | CDHOC-STL | Baseline (SVR) | CDHOC-STD | CDHOC-STL |
| 22 | 1 | 37.06% | 52.99% | 43.98% | 66.46% | 65.21% | 59.19% |
| 21 | 2 | 46.14% | 51.84% | 50.97% | 69.64% | 72.25% | 66.73% |
| 20 | 3 | 42.06% | 51.19% | 50.15% | 77.34% | 77.80% | 73.99% |
| 19 | 4 | 43.29% | 48.92% | 51.53% | 76.68% | 77.73% | 74.32% |
| 18 | 5 | 39.78% | 46.82% | 50.54% | 75.87% | 77.18% | 73.89% |
| 17 | 6 | 39.98% | 47.27% | 49.42% | 76.54% | 76.13% | 74.53% |
| 16 | 7 | 36.70% | 46.08% | 47.40% | 73.46% | 74.23% | 72 .62% |
| 15 | 8 | 35.47% | 45.02% | 44.63% | 71.83% | 72.84% | 69.12% |
| 14 | 9 | 36.01% | 45.17% | 42.51% | 70.10% | 72.10% | 69.18% |
| 13 | 10 | 37.05% | 44.41% | 43.10% | 70.26% | 72.00% | 69.16% |
| 12 | 11 | 39.16% | 46.09% | 44.31% | 71.52% | 73.85% | 70.42% |
| Average accuracy | | 39.34% | 47.80% | 46.32% | 72.70% | 73.76% | 68.92% |

Because the maximum number of people allowed in the audience at the cinema is 300, predicting the exact number of people at one particular time is challenging. The accuracy for the baseline method SVR on average is 39.4% and for CD-HOC on average is 47.8%. We decided to calculate the accuracy number with ten units error tolerance, so the difference up to ten occupants is counted as true positive. Ten units error tolerance is acceptable for cinema theatre as 280 people and 290 people will not make a huge difference for analysis and controlling HVAC/BMS purpose. The result for ten units error tolerance is shown in Figure 14. CDHOC-STL performs worse than the baseline but CDHOC-STD is the most accurate for almost every test case with an average accuracy of 73.76%. The average accuracy for SVR is 72.7%.

The results from both Figures 13 and 14 show that overall CD-HOC method is more accurate in predicting indoor human occupancy. CDHOC-STL is slightly more accurate on zero units error tolerance and CDHOC-STD is more accurate on average. This result is encouraging as CDHOC-STD uses moving average and a smoothing method usually work best with some error tolerance. Furthermore, we can observe that the accuracy for less number of days prediction is higher than for more days prediction, which is aligned with the results from academic staff room experiment.

### *5.3. Discussion*

Our research with our new framework has a high accuracy (94.4%) for a small room with up to four residents and it performs better than the baseline method for a large room with up to 300 occupants in the cinema theatre. This CD-HOC model is robust enough to handle different scales of data, proven by doing research in two different environment content such as room size and the maximum number of occupants.

From the cinema theatre result, we can observe that the CD-HOC model performs better than the baseline method and with ten units error tolerance for up to 300 occupants, the occupancy prediction accuracy is 77.8%. Estimating the number of occupants for a large room is a challenging problem and to the best of the authors' knowledge, there is no research for this large number of occupants that is generated with this level of accuracy for occupancy prediction. The best state-of-the-art performance using $CO_2$ data is 15% accuracy for predicting up to 40 occupants reported by [5].

From the experimental results, after predicting up to 10 days ahead the CD-HOC model's accuracy is reduced to only 6.9% on the average compared to predicting only one day ahead. This result is also encouraging because it shows that CD-HOC model is significantly stable for long term predictions.

The CDHOC-STL method has a higher accuracy compared to the baseline for precise accuracy but the standard accuracy decreased when we consider ten units error tolerance. This means that the value estimated can be very accurate at one time and miss the target in other cases. The ranges fluctuated significantly. CDHOC-STD, on the other hand, is more stable and has a stable range of prediction. This is aligned because STD is based on the moving average smoothing method.

## 6. Conclusions

Research on building and room occupancy counting is becoming more important. By understanding and knowing the numbers of people within a building, heating, cooling, lighting control, building energy consumption, emergency evacuation, security monitoring and room utilisation all can be made more efficient. Thorough research in this area has been studied with various methods including the use of ambient sensors. However, occupancy models that have been studied in previous work require the use of many sensors, which is expensive for installation and on-going operation. In this experiment, we use a single sensor that is commonly available in the BMS to reduce the cost and complexity as more sensors mean less reliability. The research of using a single type of information such as $CO_2$ and inferring it to predict human occupancy is novel. Hence, many possibilities can be explored by using this technique. Furthermore, with $CO_2$ data, the privacy of every single individual is protected as no personal information is required. This method is device-free in a notion that no device would be attached to the body throughout the experiment phases. Our method produces average 8.46% better accuracy compared to the baseline method. In addition, our prediction model reduces 6.9% accuracy after predicting more than 10 days.

As our research was focussed on two locations and datasets, we plan to extend this research to other places that have different environment dynamics and characteristics. The experiment above was conducted off-line. For future work, real-time on-line learning can be pursued to enhance the performance.

## Acknowledgment

The authors would like to thank Joerg Wicker from University of Mainz for providing the cinema dataset used in this paper. This research is supported by two RMIT and Siemens Sustainable Urban Precinct Project (SUPP) grants: ”iCo2mmunity: Personal and Community Monitoring for University-wide Engagement towards Greener, Healthier, and more Productive Living” and ”The Greener Office and Classroom”.

## Appendix A. Machine Learning Techniques and Their Abbreviations

Table A.1: Machine Learning Techniques and their abbreviations that are used in related works

| Abbreviation | Machine Learning Techniques |
|---|---|
| ARIMA | Autoregressive integrated moving average |
| ANN | Artificial Neural Network |
| BN | Belief Network |
| CART | Classification and Regression Trees |
| DB-SCAN | Density-Based Spatial Clustering of Applications with Noise |
| DT | Decision Trees |
| EV | Ensemble Voting |
| ELSR | Ensemble Least Square Regression |
| GBM | Gradient Boosting Machines |
| GP | Gaussian Processes |
| HMM | Hidden Markov Model |
| KNN | K-Nearest Neighbour |
| KL divergence | KullbackLeibler Divergence |
| LBMPC | Learning-Based Model Predictibe Control |
| LD | Linear Discriminant |
| LDA | Latent Dirichlet Allocation |
| LMV | Linear Minimum Variance |
| LP | Linear Regression |
| MLE | Maximum Likelihood Estimate |
| MLP | Multi-Layer Perceptron |
| NB | Na¨ıve Bayes |
| NMF | Non-negative Matrix Factorization |
| NN | Neural Networks |
| MP | MultiPath fading |
| RF | Random Forest |
| RBF | Radial Basis Function |
| SVM | Support Vector Machine |
| SVR | Support Vector Regression |
| TAN | Tree Augmented Na¨ıve Bayes network |
| THR | Thresholding |